\documentclass[aps,prb,reprint]{revtex4-1}
\usepackage[dvipdfm]{graphicx}
\usepackage{amsmath}

 \draft 

\usepackage{bm}
\def\vb#1{{\bm#1}}
\def\v#1{\mathbf{#1}}		
\def\r{\v{r}} 					
\def\p{\v{p}} 					
\def\k{\v{k}} 					

\def\vOmega{\vb{\Omega}}
\def\vpi{\vb{\pi}}
\def\vsigma{\vb{\sigma}}
\def\valpha{\vb{\alpha}}
\def\vSigma{\vb{\Sigma}}
\def\vbeta{\vb{\beta}}

\def\B{\v{B}}
\def\E{\v{E}}

\def\vv{\v{v}}
\def\vI{\v{v}_{\rm I}}

\def\va{\v{a}}
\def\ve{\v{e}}

\def\mat#1#2#3#4{\left(
\begin{array}{cc} #1 & #2 \\#3& #4\end{array} \right)}
\def\del{\partial}

\def\CE{\mathcal{E}}
\def\CO{\mathcal{O}}

\def\CF{\mathcal{F}}

\def\bgamma{\bar{\gamma}}

\begin{document}


\title{Spin-dependent inertial force and spin current in accelerating systems} 



\author{Mamoru Matsuo$^{1,2}$,
Jun'ichi Ieda$^{2,3}$, Eiji Saitoh$^{2,3,4}$,
and Sadamichi Maekawa$^{2,3}$ }
\affiliation{%
$^{1}$Yukawa Institute for Theoretical Physics,  Kyoto University, Kyoto 606-8502, Japan \\
$^{2}$The Advanced Science Research Center, Japan Atomic Energy Agency, Tokai 319-1195, Japan \\
$^{3}$CREST, Japan Science and Technology Agency, Sanbancho, Tokyo 102-0075, Japan\\
$^{4}$Institute for Materials Research, Tohoku University, Sendai 980-8577, Japan}

\date{\today}

\begin{abstract}
The spin-dependent inertial force in an accelerating system under the presence of electromagnetic fields is derived from the generally covariant Dirac equation. 
Spin currents are evaluated by the force up to the lowest order of the spin-orbit coupling in both ballistic and diffusive regimes. 
We give an interpretation of the inertial effect of linear acceleration on an electron as an effective electric field 
and show that mechanical vibration in a high frequency resonator can create a spin current via the spin-orbit interaction augmented by the linear acceleration.
\end{abstract}


\maketitle 

\section{Introduction}
Studies of inertial effects on electrons have a long history dating back to 1910s.
Barnett investigated the magnetization induced by rotational motion.\cite{Barnett1915}
 Einstein and de Haas carried out the reversed experiment.\cite{Einstein-deHaas1915}
They measured the gyromagnetic ratio and the anomalous $g$ factor of electrons before the establishment of modern quantum physics.
Stewart and Tolman estimated the electron mass by measuring the charge accumulation at the rim of a metal due to the linearly accelerating motion.\cite{Tolman1916}
Rapid progress in nanotechnology has allowed us to study the coupling of mechanical motion and electromagnetism in the quantum mechanical regime.
Effects of mechanical rotation on nanostructured magnetism are detected in 
microcantilevers\cite{Rugar1992,Wallis2006} and 
a torsional resonator\cite{Zolfagharkhani2008}.
The quantization of the rotational motion is observed in magnetic nanoparticles.\cite{Tejada2010}
Single quantum excitations of vibration, namely, phonons, can be controlled in a piezoelectric acoustic wave resonator.\cite{Connell2010}
There has been theoretical work on 
 effects of mechanical rotation on nanostructured magnetism,\cite{Mohanty2004,Kovalev2007,Bretzel2009,Jaafar2009prb,Chudnovsky2010,Lendinez2010,Bauer2010} the coupling of nanomechanical vibration and magnet,\cite{Kovalev2003,Kovalev2005,Jaafar2009prl,Kovalev2011} and the phonon-spin coupling related to spin relaxation.\cite{Chudnovsky2002,Chudnovsky2004,Chudnovsky2005,Calero2005,Calero2007prl,Calero2007prb,Kim2009}
However, the contribution of the spin-orbit interaction (SOI) in accelerating frames has not thoroughly been studied in the previous papers.

Recent developments in spintronics,\cite{MaekawaEd2006,Bader2010} which relies on not only electron's charge but also its spin, have enabled us to utilize a ``spin current'', a flow of spins.
In this context, the coupling of magnetization and spin current  is of great interest in the field of spintronics.\cite{Slonczewski1996,Berger1996,Tserkovnyak2002,Barnes2007}
To harness the spin current, the understanding of the spin-orbit interaction is indispensable.
Recently, the present authors proposed 
a fundamental theory describing the direct coupling of the mechanical rotation and spin current and predicted the spin current generation arising from rotational motion.\cite{Mamoru2011,Mamoru2011a}
Our finding has made the first step to extend the theory of spin current in the inertial frame to that in the noninertial frame.
In this paper, we provide a systematic approach to study spin current generation from mechanical motion, including time-dependent rigid rotation and linear acceleration. 
Firstly, we derive the spin-dependent inertial force in a rotating frame in the presence of an applied magnetic field. This force is responsible for the generation of a spin current due to mechanical rotation.
Secondly, generation of the spin current by linear acceleration is investigated. We show that nanomechanical vibration can create an ac spin current on the basis of the inertial spin-orbit coupling in accelerating systems.

The outline of the paper is the following. In Sec. \ref{Sec:DiracEquation}, we review the Dirac equation in a rotating frame. 
In Sec. \ref{Sec:PauliRot}, the Pauli-Schr\"odinger equation in the rotating frame is derived.
In Sec. \ref{Sec:SpinDependentForce}, we derive a full expression of a spin-dependent force caused by mechanical rotation.  
Spin current generation in the presence of impurity scattering is studied in Sec. \ref{Sec:Impurity}. 
A detection method of such spin currents is proposed in Sec. \ref{Sec:DetectISHE}.
Spin current generation from mechanical vibration is investigated in Sec. \ref{Sec:Linear}. 
The paper ends with a few concluding remarks in Sec. \ref{Sec:Conclusion} and three Appendices.
Appendix \ref{Sec:Metric} contains short summaries on vierbein. Details of Foldy-Wouthuysen-Tani transformation are given in Appendix \ref{Sec:FWTT}. 
Electromagnetic fields in a rotating frame are briefly summarized in Appendix \ref{Sec:EMrot}. 
The spin diffusion equation in a rotating frame is solved in Appendix \ref{Sec:SpinDiffusionEquation}.

\section{Dirac equation in a rotating frame\label{Sec:DiracEquation}}
In this section, we review the Dirac equation in a rotating frame. 
According to Einstein's principle of equivalence, gravitation cannot be locally distinguished from inertial effects due to acceleration of the frame of references.
In the general relativity, both gravitational and inertial effects are expressed by a metric and connection in a curved space-time.
The fundamental equation of a spin-1/2 particle in a curved space-time is the generally covariant Dirac equation\cite{Bib:SpinConnection}:
\begin{eqnarray}
\left[\gamma^{\mu}\left( \del_{\mu }- \Gamma_{\mu} - \frac{iqA_{\mu}}{\hbar}   \right) + \frac{mc}{\hbar}  \right] \Psi =0,\label{gDirac}
\end{eqnarray}
where $c, \hbar, q=-e,$ and $ m$ are the velocity of light, the Planck constant, the charge and mass of an electron,
$A_{\mu}=(A_{0},\v{A})$ is the U(1) gauge potential, and $\Gamma_{\mu}$ the spin connection\cite{Bib:SpinConnection}.
The Clifford algebra in the curved space-time $\gamma^{\mu}=\gamma^{\mu}(x)$ satisfies 
\begin{eqnarray}
\{ \gamma^{\mu}(x), \gamma^{\nu}(x) \}= 2g^{\mu\nu}(x)
\end{eqnarray}
where $g^{\mu\nu}(x)\ (\mu, \nu = 0,1,2,3)$ is the inverse of the coordinate-dependent metric $g_{\mu\nu}(x)$.
The coordinate transformation from 
a rigidly rotating frame to the inertial frame is given by
\begin{eqnarray}
d\v{r}'=d\v{r} + (\vb{\Omega} \times \v{r}) dt,
\end{eqnarray}
where the rotation frequency with respect to an inertial frame is $\vb{\Omega}(t)$.
Here we assume that the rotation velocity $\vOmega \times \r$ is much less than the speed of light.
The space-time line element in the rotating frame is
\begin{eqnarray}
ds^{2}&=&g_{\mu\nu}dx^{\mu}dx^{\nu} \nonumber\\
&=&[-c^{2} + (\vb{\Omega} \times \r )^{2}] dt^{2} + 2 (\vb{\Omega} \times \r)dt d\r +d\r^{2}.\label{ds2}
\end{eqnarray}
Thus, the metric in the rotating frame becomes 
\begin{eqnarray}
g_{\mu \nu} =
\left(\begin{array}{cccc}
-1+\v{u}(x)^2 & u_x(x) & u_y(x) & u_z(x) \\
u_x(x) & 1 & 0 & 0 \\
u_y(x) & 0 & 1 & 0 \\
u_{z}(x) & 0 & 0 & 1\end{array}\right),
\end{eqnarray}
with
\begin{eqnarray}
\v{u}(x)= \vb{\Omega}(t) \times \r/c.
\end{eqnarray}
This metric leads to the Clifford algebra and the spin connection in a rotating frame as
\begin{eqnarray}
\gamma^{0}(x)&=& i\beta, \ \gamma^{i}(x)=i\beta \alpha_{i}- u_{i}(x),\\
\Gamma_{0}&=& \frac{\vb{\Omega}\cdot \vb{\Sigma}}{2c}, \ \Gamma_{i}=0,\label{SpinConnection}
\end{eqnarray}
where 
\begin{eqnarray}
 \beta=\mat{I}{O}{O}{-I}, \ 
 \vb{\alpha}=\mat{O}{\vb{\sigma}}{\vb{\sigma}}{O}
\end{eqnarray}
 are the Dirac matrices and $\vb{\Sigma} $ is the spin operator for 4-spinor
defined by 
\begin{eqnarray}
\vb{\Sigma} =\frac{\hbar }{4i} \vb{\alpha} \times \vb{\alpha} = \frac{\hbar}{2} \mat{\vb{\sigma}}{O}{O}{\vb{\sigma}}
\end{eqnarray} with  
the Pauli matrix $\vb{\sigma}$ (details of the spin connection are given in Appendix \ref{Sec:Metric}).
From Eqs. (\ref{gDirac})-(\ref{SpinConnection}), the Dirac equation in a rotating frame is written as
\begin{subequations}
\begin{eqnarray}
&&i\hbar \frac{\del \Psi}{\del t}=H \Psi,  \\
&& H = \beta mc^{2} + c \vb{\alpha} \cdot  \vb{\pi} +qA_{0}
	-\vb{\Omega}\cdot \left( \v{r}\times \vb{\pi}+\vb{\Sigma} \right), \label{Hrot}
\end{eqnarray}
\end{subequations}
where   
$\vb{\pi}=\p - q\v{A} $ is the mechanical momentum and $\r$ is the position vector from the origin at the rotation axis.

In classical mechanics, the Hamiltonian in the rotating frame has the additional term $\vb{\Omega} \cdot (\r \times \vb{\pi})$ which reproduces the inertial effects: Coriolis, centrifugal, and Euler forces\cite{LandauMech}.
The term $\vb{\Omega}\cdot \vb{\Sigma}$ is called the spin-rotation coupling.\cite{Oliveira1962,Mashhoon1988,Hehl1990}
The last term of Eq. (\ref{Hrot}), $\vb{\Omega}\cdot (\r \times \vb{\pi} + \vb{\Sigma})$, can be regarded as
a quantum mechanical generalization of the inertial effects 
obtained by replacing the mechanical angular momentum $\r \times \vb{\pi}$ with the total angular momentum $\r \times \vb{\pi} + \vb{\Sigma}$.

\section{Pauli--Schr\"odinger equation in a rotating frame\label{Sec:PauliRot}}
The Dirac equation is an equation of 4-spinor wave function, which contains the up/down-spin electron and positron components.
As the energy gap between the electron and positron state is much larger than the energy level of condensed matter systems,
we take the low energy limit of the Dirac equation to obtain the Pauli-Schr\"odinger equation of up- and down-spin electrons.
Following the low energy expansion and block-diagonalization method of the Dirac Hamiltonian developed by Foldy, Wouthuysen,\cite{Foldy1950} and Tani,\cite{Tani1951}
we derive the Pauli-Schr\"odinger equation in a rotating frame (see Appendix \ref{Sec:FWTT} for the detail of the derivation).
The Hamiltonian (\ref{Hrot}) is divided into the block diagonal and off-diagonal parts denoted by $\CE$ and $\CO$, respectively;
\begin{eqnarray}
H&=& \beta mc^{2} + \CE + \CO,
\end{eqnarray}
with
\begin{subequations}
\begin{eqnarray}
\CE &=& qA_{0} - \vb{\Omega} \cdot (\r \times \vb{\pi}+ \vb{\Sigma}),  \\
\CO &=& c\vb{\alpha} \cdot \vpi. 
\end{eqnarray}
\end{subequations}
By successive Foldy-Wouthuysen-Tani transformations, the Hamiltonian up to the order of $1/m^{2}$ becomes
\begin{eqnarray}
H =&& \beta \left[ mc^{2}+ \frac{\CO^{2} }{2mc^{2}}  \right] +\CE  
- \frac{1}{8m^{2}c^{4}} \left[\CO, \left[ \CO, \CE \right]+ i\hbar \dot{\CO} \right]. \nonumber \\ \label{Hfwt}
\end{eqnarray}
Neglecting the rest energy term in Eq. (\ref{Hfwt}), the Pauli-Schr\"odinger equation for the upper component of Dirac spinors, namely, 2-component electron wave function, in the rotating frame is obtained as\cite{Mamoru2011}
\begin{eqnarray}
&& i\hbar \frac{\del \psi}{\del t}  = H_{{\rm PR}} \psi, \label{PReq} \\
&&H_{{\rm PR}}=H_{{\rm K}}+H_{{\rm Z}}+H_{{\rm I}}+H_{{\rm S}}+H_{{\rm D}}, \label{Hpr}
\end{eqnarray}
where
\begin{subequations}
\begin{eqnarray}
&&H_{{\rm K}}=  \frac{1}{2m}  \vb{\pi}^{2} + qA_{0}, \\
&&H_{{\rm Z}}= \mu_{B} \vb{\sigma} \cdot \v{B} ,\\
&&H_{{\rm I}}=- \vb{\Omega}\cdot (\r \times \vb{\pi}+\v{S}), \label{HI}  \\
&&H_{{\rm S}}=\frac{q\lambda}{2\hbar} \vb{\sigma}\cdot ( \vb{\pi} \times  \v{E}'-\v{E}' \times \vb{\pi}), \label{HS} \\
&&H_{{\rm D}}= -\frac{q\lambda}{2} \mbox{div} \v{E}',
\end{eqnarray}
\end{subequations}
with 
\begin{eqnarray}
 \mu_{B}=\frac{q\hbar}{2m}, \ \lambda =\frac{\hbar^{2}}{4m^{2}c^{2}}, \
 \v{S}=\frac{\hbar}{2}\vb{\sigma},
\end{eqnarray}
 and
\begin{eqnarray}
\v{E}'= \v{E} + ( \vb{\Omega}\times \r) \times \v{B}.  \label{Erot}
\end{eqnarray}

The Hamiltonian in Eq. (\ref{PReq}), $H_{{\rm PR}}$, is a $2 \times 2$ matrix operator and $\psi$ is the 2-spinor wave function of a single electron.
\subsection{Lowest order of the expansion}
In the lowest order of the expansion, the Hamiltonian to the order of $1/m$ is given by $H_{{\rm K}}+H_{{\rm Z}}+H_{{\rm I}}$.
The spin-independent $H_{{\rm K}}$ contains the kinetic energy and the potential energy.
The Zeeman energy $H_{{\rm Z}}$ contains the $g$ factor of the electron equal to 2.
Combined $H_{{\rm K}}$ with $H_{{\rm Z}}$, the coupling with magnetic field, 
\begin{eqnarray}
\frac{q}{2m}(\r \times \vb{\pi} + 2 \v{S})\cdot \B
\end{eqnarray}
 is obtained,
which contrasts with Eq. (\ref{HI}): the mechanical rotation couples to the total angular momentum of the electron
\begin{eqnarray}
\r \times \vb{\pi} + \v{S}.
\end{eqnarray}
The inertial effects, namely, the Coriolis, centrifugal, and Euler forces, are reproduced by the first term of $H_{I}$ as mentioned above. 
The second term of $H_{{\rm I}}$ is the spin rotation coupling term.
Introducing the ``Barnett field'', 
\begin{eqnarray}
\v{B}_{\vOmega}=(m/q)\vb{\Omega}, 
\end{eqnarray}
we can combine the spin rotation coupling with the Zeeman term, leading to a different form:
\begin{eqnarray}
\mu_{B}\vb{\sigma}\cdot (\v{B} + \v{B}_{\vOmega}), \label{brot} 
\end{eqnarray}
Equation (\ref{brot}) shows that the spin-rotation coupling can be interpreted as
a correction to the Zeeman effect with an effective magnetic field $\v{B}_{\vOmega}$.

Previous theoretical work\cite{Mohanty2004,Kovalev2007,Bretzel2009,Jaafar2009prb,Chudnovsky2010,Lendinez2010,Bauer2010} has been done on the basis of the Barnett field. 
In the following sections, we study inertial effects on spin current using the spin-orbit interaction $H_{\rm S}$ which is obtained from the second order of the expansion.

\subsection{Second order of the expansion\label{SubSec:2nd}}
The expansion of the order of $1/m^{2}$ yields the SOI and Darwin term with the mechanical rotation, $H_{\rm S}$ and $H_{\rm D}$. 
In the absence of the rotation, $\vb{\Omega}=\v{0}$, these terms reproduce the conventional SOI and Darwin terms
in the rest frame. In the presence of the rotation, we find that the electric field $\v{E}$ in the two terms in the inertial frame
is modified by an additional term $(\vb{\Omega} \times \r) \times \v{B}$.
This result is consistent with a general coordinate transformation of electromagnetic fields between the rest frame and the rotating frame\cite{Ridgely1999} (the derivation of the transformation is given in Appendix \ref{Sec:EMrot}).

\subsection{Renormalization of SOI}
The contribution of $H_{\rm S}$ to $H_{\rm I}$ in vacuum is negligible since the dimensionless spin-orbit coupling parameter
\begin{eqnarray}
\eta_{\rm SO}=\frac{\lambda (mv)^{2}}{ \hbar^{2}} = \left( \frac{v}{2c} \right)^{2} \ll 1
\end{eqnarray}
with the electron velocity $v$.
However, the spin-orbit coupling is enhanced in metals and semiconductors 
such as Pt.\cite{Vila2007,Takahashi2008}
The renormalization depends on detailed electronic structures and electron correlations.\cite{Guo2009,Gu2010}
In the present study, we do not go into detail about this procedure. Nevertheless, the results obtained in this paper are universal in nature and one can start with an effective Hamiltonian, e.g., Luttinger\cite{Murakami2003} or Rashba\cite{Sinova2004} model, which refers the electromagnetic fields (\ref{Erot}) in a rotating frame.
Replacing the momentum $mv$ with the Fermi momentum $\hbar k_{F}$, the coupling $\eta_{\rm SO}$ becomes $\tilde{\lambda}k_{F}^{2}$ where $\tilde{\lambda}$ is an enhanced spin-orbit coupling parameter. 
The coupling $\eta_{\rm SO}$ of Pt is estimated as 0.59 by the nonlocal measurement of the spin Hall effect.\cite{Vila2007,Takahashi2008}
Electrons in a noninertial frame cannot distinguish the inertial effect originating from the mechanical rotation $(\vb{\Omega} \times \r) \times \v{B}$ in Eq. (\ref{HS}) from conventional electric field $\v{E}$. 
Thus, the new effect due to the SOI with the mechanical rotation can be sizable effects in the large SOI systems as shown in the following sections.

\section{spin-dependent inertial force in a rotating frame\label{Sec:SpinDependentForce}}
Let us consider semi-classical equations of motion for an electron based on the Pauli-Schr\"odinger equation in a rotating frame.
A quantum mechanical analogue of  a ``force'' $\CF$ is defined by 
\begin{eqnarray}
\CF=\frac{1}{i\hbar}[m\dot{\r},H_{\rm PR}] + m\frac{\del \dot{\r}}{\del t}, \label{SpinForce}
\end{eqnarray}
with $\dot{\r}=[\r,H_{\rm PR}]/i\hbar$.
From Eq. (\ref{PReq}), the spin-dependent velocity including the effect of mechanical rotation is obtained as 
\begin{eqnarray}
&&\dot{\v{r}}=\v{v} + \v{v}_{\rm I} + \v{v}_{\vb{\sigma}},\label{dotR-HPR}
\end{eqnarray}
with
\begin{subequations}
\begin{eqnarray}
&&\v{v}=\frac{1}{i\hbar}[\r,H_{\rm K}]=\frac{\vb{\pi}}{m}, \\
&&\v{v}_{\rm I}=\frac{1}{i\hbar}[\r,H_{\rm I}]=- \vb{\Omega}\times \r, \\
&& \v{v}_{\vb{\sigma}}=\frac{1}{i\hbar}[\r,H_{\rm S}] = \frac{e\lambda}{\hbar} \vb{\sigma}\times \v{E}'.
\end{eqnarray}
\end{subequations}
The ``force'' $\CF=\CF_{0}+\CF_{1}+ \CF_{2}+\CF_{t}$ is obtained as
\begin{subequations}
\begin{eqnarray}
\CF_{0}=&&q [\v{E}' + \vv \times (\v{B} + 2\B_{\Omega})]  + m \vOmega \times (\vOmega \times \r), \\
\CF_{1}=
&& -\frac{q^{2}\lambda}{\hbar} \{  (\vsigma \times \E') \times  (\B+\B_{\Omega})-[(\B + \B_{\Omega})\times \vsigma] \times \E' \} \nonumber \\
&& + \frac{qm\lambda}{\hbar} [  \vsigma \cdot (\vOmega \times  \vv) \v{B}  +2 (\B\cdot \vOmega) \vsigma \times \vv   - (\B \cdot \vv)\vsigma \times \vOmega  \nonumber\\
&&+ \vOmega \cdot (\r \times \B) \vsigma \times \vOmega -(\B \cdot \vOmega) \vsigma\times (\vOmega \times \r)], \\
\CF_{2}=&& \frac{mq^{2} \lambda^{2}}{\hbar^{2}} \Big[ \frac2{\hbar}(\vsigma \cdot \E') m \vv \times \E'  \nonumber\\
&&+ i [ (\vsigma \cdot \E')\B \times \vOmega - (\vsigma \cdot \B) \E' \times \vOmega  + (\B \cdot \vOmega) \E' \times \vsigma  ] \nonumber\\
&& + (\E' \times \B) \times \vOmega + 2 (\B \cdot \vOmega) \E' \Big],\\
\CF_{t}=&&  m\r \times \frac{\del \vOmega}{\del t}  + \frac{qm\lambda}{\hbar}\vsigma\times \left[ \left(\r \times \frac{\del \vOmega }{\del t} \right) \times \B \right].
\end{eqnarray}
\end{subequations}
The spin-independent $\CF_{0}$ consists of the electromagnetic force in a rotating frame $q (\v{E}' + \vv \times \v{B})$, Colioris force $q\vv \times  2\B_{\Omega}=2 m\vv \times \vOmega$, and centrifugal force $m \vOmega \times (\vOmega \times \r)$.
The first term in $\CF_{t}$ is the Euler force. 
The other terms in $\CF$ with the spin operator $\vsigma$ are responsible for spin-dependent transport of electrons.
The full expression of the spin-dependent force in a rotating frame in the presence of electromagnetic fields is one of the principal results of this paper.
In the absence of rotation, $\vOmega=\v{0}$, the above expression of $\CF$ reproduces the previous results in an inertial frame.\cite{Shen2005} 

In order to have the better understanding of the spin-dependent force, we study the case of $\vOmega=(0,0,\Omega), \B=(0,0,B), \E=\v{0}, |\B/\B_{\Omega}| \gg 1$ and neglect the terms of the order of $\eta_{\rm SO}^{2}, |\Omega/\omega_{c}|^{2}$ with the cyclotron frequency 
\begin{eqnarray}
\omega_{c}=\frac{qB}{m}.
\end{eqnarray}
$\CF$ is decomposed into the $xy$- and $z$-components,  $\CF_{\perp}$ and $\CF_{\parallel}$. 
Thus, we have
\begin{eqnarray}
&&\CF_{\perp} \approx q (\E_{\r} + \E_{\vsigma} + \vv \times \B), \label{CFperp}
\end{eqnarray}
with
\begin{subequations}
\begin{eqnarray}
&&\E_{\r}  = (\B\cdot \vOmega)\r,  \\
&&\E_{\vsigma}=- \frac{q\lambda}{\hbar}(\B \cdot \vsigma)(\B \cdot \vOmega)\r. \label{ErEspin}
\end{eqnarray} 
\end{subequations}
Here, $\E_{\r}$ is an electric field induced in the rotating frame with an applied magnetic field $\B$, and $\E_{\vsigma}$ is an ``effective spin-dependent electric field'' induced by the SOI, $H_{S}$. 
Coexistence of electric and magnetic fields, $\E$ and $\B$, yields the $\E \times \B$ drift, which is the motion of guiding center of a charged particle with the drift velocity\cite{LandauFields}
\begin{eqnarray}
 \vv^{\rm d}= \frac{\E \times \B}{B^{2}}.
\end{eqnarray}
From Eq. (\ref{CFperp}) we obtain two types of the drift motion for an electron wave packet: 
one is the charge drift motion with 
\begin{eqnarray}
\vv_{c}^{\rm d}=\frac{\E_{\r} \times \B}{B^{2}},
\end{eqnarray}
and 
the other is the spin-dependent drift motion with
\begin{eqnarray}
\vv_{\vsigma}^{\rm d}=\frac{\E_{\vsigma} \times \B}{B^{2}}.
\end{eqnarray} 
Figure 1 (a) illustrates the relation of the rotation, magnetic field, induced spin-dependent field, and drift velocity.
In a ballistic regime, the latter produces the spin current in the azimuthal direction, 
\begin{eqnarray}
\v{J}_{s} = en{\rm Tr} \sigma_{z} \vv^{\rm d}_{\vsigma}=2ne\kappa \omega_{c}R \v{e}_{\phi},\label{JsDrift}
\end{eqnarray} 
where $R$ is the distance from the rotation axis, $\v{e}_{\phi}$ the unit azimuthal vector, $n$ the electron density, and the dimensionless parameter 
\begin{eqnarray}
\kappa = \tilde{\lambda} k_{F}^{2}\cdot \frac{\hbar \Omega}{\epsilon_{F}}
\end{eqnarray}
with Fermi energy $\epsilon_{F}$.\cite{Mamoru2011}
Setting $B=1$T, $\Omega=1$kH, $\tilde{\lambda} k_{F}^{2}\approx 0.6$, $k_{F} \approx 10^{10}$m, and $R=10$mm, 
$|\v{J}_{s}|$ is estimated to be about $10^{8}{\rm A/m^{2}}$.

\begin{figure}[tbp]
\begin{center}
\includegraphics[scale=0.5]{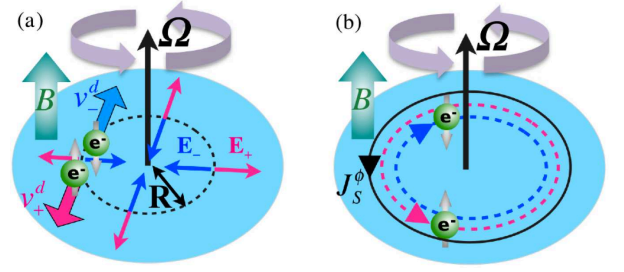}
\end{center}
\caption{(a) Spin-dependent electric field  $\E_{\vsigma}$ and drift velocity $\vv^{\rm d}_{\vsigma}$ are illustrated. 
An external magnetic field $\B$ is applied along the rotation axis ($z$-direction). For the $z$-polarized spins, the electric field, $\E_{+}$($\E_{-}$), is induced in the radial outward(inward) direction. (b) The drift velocities $\vv^{\rm d}_{\pm}$ in opposite directions result in the spin current $J_{s}^{\phi}$ in the azimuthal direction. }
\label{setup}
\end{figure}

Next, we consider fluctuation of the rotating axis, caused by a high speed rotor in operation.
Such fluctuation effects on spin current come from $\vOmega$ dependence of $\E_{\vsigma}$ in Eq. (\ref{ErEspin}) as well as the time-derivative of the $\vOmega$ in the second term of $\CF_{t}$.
If the rotation frequency has a time-dependent component as
\begin{eqnarray}
\vOmega (t) = \vOmega_{0} + \delta \vOmega (t)
\end{eqnarray}
where 
\begin{eqnarray}
\delta \vOmega (t) = \delta \vOmega_{\perp} (t)+ \delta \vOmega_{\parallel} (t),\label{deltavOmega}
\end{eqnarray}
with $\delta \vOmega_{\perp} \perp \vOmega_{0}$, $\delta \vOmega_{\parallel} \parallel \vOmega_{0}$, and $|\delta \vOmega | \ll |\vOmega|$,
time-dependent drift motion is induced:
\begin{eqnarray}
 \delta \vv_{\vsigma}^{\rm d} = \frac{\delta \E_{\vsigma} \times \B}{B^{2}}
\end{eqnarray}
with 
\begin{eqnarray}
\delta \E_{\vsigma}= -\frac{q\lambda}{\hbar} (\vsigma \cdot \B) (\B \cdot \delta\vOmega)\r,
\end{eqnarray}
and the fluctuation of the spin current is obtaind as 
\begin{eqnarray}
\delta\v{J} _{s}(t)=\frac{2ne^{2} \lambda BR }{\hbar} \delta \vOmega_{\parallel}(t) \v{e}_{\phi}.
\end{eqnarray}
The time-derivative of Eq. (\ref{deltavOmega}) is divided into the $xy$- and $z$-components: 
\begin{eqnarray}
\frac{\del \delta \vOmega}{\del t} = \left( \frac{\del \delta \vOmega}{\del t} \right)_{\perp} + \left( \frac{\del \delta \vOmega}{\del t} \right)_{\parallel}.
\end{eqnarray}
From the second term of $\CF_{t}$, we have an additional effective spin-dependent electric field created in the $xy$-plane $\delta \E_{\vsigma}'$ as
\begin{eqnarray}
\delta \E_{\vsigma}'(t) =\frac{qm\lambda}{\hbar} (\vsigma \cdot \B) \r \times \left( \frac{\del \delta\vOmega}{\del t} \right)_{\parallel} .
\end{eqnarray}
This yields spin current in the azimuthal direction 
\begin{eqnarray}
\delta \v{J}'_{s}(t) = \frac{2nem \lambda R}{\hbar} \left( \frac{\del \delta\vOmega}{\del t} \right)_{\parallel}  \v{e}_{\phi}. 
\end{eqnarray}
The ratio $|\delta J_{s}'/\delta J_{s} |$ is equal to $\left|\frac{\del \delta \Omega}{\del t} \right|/|\delta \Omega \omega_{c}|$. 
The time scale of the fluctuation of rotation is usually much smaller than that of the cyclotron frequency. 
Thus, the contribution from $\CF_{t}$ to the spin current fluctuation is negligible.

\section{effects of impurity scattering\label{Sec:Impurity}}
In this section, we discuss the spin current generation in the presence of impurity scattering. 
We consider a Pt thin film attached on a rotating disk with a uniform magnetic field parallel with the rotation axis as shown in Fig. \ref{setup}.
Spin dependent transport in a system with a strong SOI can be described by semiclassical equations\cite{Xiao2010}
\begin{subequations}\label{SemiClassical}
\begin{eqnarray}
\dot{\v{r}} &=& \v{v} + \v{v}_{\vb{\sigma}0},  \label{SemiClassical-a} \\
\hbar \dot{\v{k}}&=& -e \left( \E +  \dot{\v{r}} \times \v{B} \right). \label{SemiClassical-b}
\end{eqnarray}
\end{subequations}
Here, $\v{v}=\hbar\k/m$ represents the normal velocity, $\v{k}$ is the wave vector, $\v{E}$ and $\v{B}$ are applied electric and magnetic fields. 
The anomalous velocity $v_{\vsigma0}$ originating from the SOI in an inertial frame is written as
\begin{eqnarray}
\v{v}_{\vb{\sigma}0} = \frac{e\lambda}{\hbar}\vb{\sigma} \times \v{E}.
\end{eqnarray}

It is straightforward to extend the equations in an inertial frame to that in a rotating frame using Eqs. (\ref{SpinForce}) and (\ref{dotR-HPR}),
\begin{subequations}\label{SemiClassicalRot}
\begin{eqnarray}
\dot{\v{r}} &=& \v{v} +\vI+ \v{v}_{\vb{\sigma}}, \label{SemiClassicalRot-a} \\
\hbar \dot{\v{k}} &=& \CF. \label{SemiClassicalRot-b}
\end{eqnarray}
\end{subequations}
Let us consider spin current generation in a rotating normal metal with the large spin-orbit coupling such as Pt in the presence of spin-independent impurity scattering.
The electron distribution function depends on spins because of the spin-dependence of the semi-classical equations (\ref{SemiClassicalRot}). In this case, the transport equation of non-equilibrium steady states is written as
\begin{eqnarray}
\dot{\r} \cdot \frac{\del f_{\vb{\sigma}}}{\del \r} + \dot{\v{k}} \cdot \frac{\del f_{\vb{\sigma}}}{\del \v{k}}  = - \frac{ f_{\vb{\sigma}} - f_{0} }{ \tau }, \label{BoltzmannEq}
\end{eqnarray}
where $f_{\vb{\sigma}}=f_{\vb{\sigma}}(\v{r},\v{k})$ is the spin-dependent distribution function, $f_{0}=f_{0}(\varepsilon)$ the Fermi-Dirac distribution function, and $\tau$ the relaxation time.

Combining the semiclassical equations (\ref{SemiClassicalRot}) and  with Eq. (\ref{BoltzmannEq}), 
by putting  
$\v{E}=\v{0}$ and 
\begin{eqnarray}
\CF =\CF_{\perp}\approx -e[\v{E}_{\r} + \v{E}_{\vsigma}+ \vv \times \B],
\end{eqnarray}
the solution is
\begin{eqnarray}
f_{\vb{\sigma}} = f_{0} + e\v{v} \cdot   \tau \frac{ \v{E}'_{\vsigma} + \tau \vb{\omega}_{c} \times \v{E}'_{\vsigma} }{1+ ( \tau \omega_{c} )^{2}  } \frac{ \del f_{0} }{ \del \varepsilon  },\label{Sol:Boltzmann-a}
\end{eqnarray}
with 
\begin{eqnarray}
\vb{\omega}_{c} = \frac{e \v{B}}{m} \label{Sol:Boltzmann-b}
\end{eqnarray}
and
\begin{eqnarray}
\v{E}'_{\vsigma} = \E_{\r} + \E_{\vsigma}. \label{Sol:Boltzmann-c}
\end{eqnarray}
The solution (\ref{Sol:Boltzmann-a}) contains two ``electric fields'': the spin-independent part $\E_{\r}$ and the spin-dependent one $\E_{\vsigma}$ as discussed in Sec. \ref{Sec:SpinDependentForce}. 
The spin-independent part yields the conventional Hall effect in the rotating frame. If the ends in the radial (longitudinal) direction of Pt film attached to the rotating disk (see Fig. \ref{setup}) are electrically connected, the Hall voltage is obtained in the azimuthal (transverse) direction while the spin-dependent part causes the spin current generation.
\begin{figure}[tbp]
\begin{center}
\includegraphics[scale=0.55]{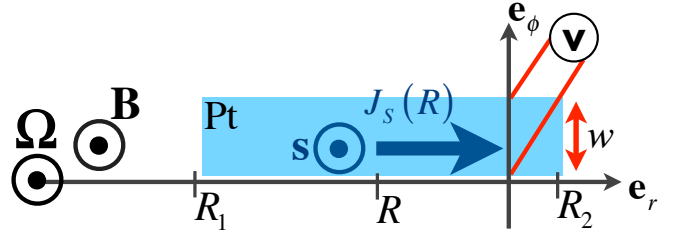}
\end{center}
\caption{The $z$-polarized radial spin current is converted to the inverse spin Hall voltage in the azimuthal direction $\v{e}_{\phi}$. The rotation frequency $\vOmega$ and the magnetic field $\v{B}$ are applied along the $z$-axis.  The $z$-polarized spin current $J_{s}(R)$ is induced in the radial direction $\v{e}_{r}$. Here, the spin polarization vector is denoted by $\v{s}$.}
\label{setup}
\end{figure}
The $z$-polarized spin current generated by the mechanical rotation can be evaluated by
\begin{eqnarray}
\v{J}_{s}=-e    \int d\k {\rm Tr}\left[\sigma_{z} f_{\vb{\sigma}}(\r,\k) \dot{\r}  \right],
\end{eqnarray}
which leads to the explicit form of the spin current:
\begin{eqnarray}
&& \v{J}_{s}(R) = J_{s}^{r}(R) \v{e}_{r} + J_{s}^{\phi}(R) \v{e}_{\phi}, \label{Js:r-phi}
\end{eqnarray}
with
\begin{subequations}
\begin{eqnarray}
&& J_{s}^{r}   =  \frac{ \tau \omega_{c} }{1+ (\tau \omega_{c})^{2} } J_{s}^{0}, \\
&&J_{s}^{\phi}  =  \frac{ (\tau \omega_{c})^{2} }{1+ (\tau \omega_{c})^{2} } J_{s}^{0}. 
\end{eqnarray}
\end{subequations}
Here, $\v{e}_{r}(\v{e}_{\phi})$ is the unit vector of the radial (azimuthal) direction as indicated in Fig. \ref{setup}, and $J_{s}^{0} $ is given by 
\begin{eqnarray}
J_{s}^{0} (R)= 2ne \kappa \omega_{c} R. \label{Js0}
\end{eqnarray}
In the large $\omega_{c} \tau$ limit, 
the radial spin current, $J_{s}^{r}$, vanishes and we have 
\begin{eqnarray}
J_{s}^{\phi} \to J_{s}^{0}.
\end{eqnarray}
This reproduces the ballistic case, Eq. (\ref{JsDrift}).
Putting $\omega_{c}\tau \ll 1$, the radial spin current bacomes much larger than the azimuthal one:
\begin{eqnarray}
J_{s}^{r} \gg J_{s}^{\phi}.
\end{eqnarray}
In the following section, we consider the latter case.

\section{Detection method of the rotationally induced spin current\label{Sec:DetectISHE}}
\begin{figure}[tbp]
\begin{center}
\includegraphics[scale=0.55]{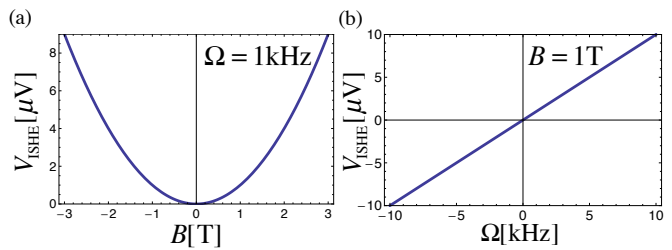}
\end{center}
\caption{The voltage induced by the inverse spin Hall effect in a rotating Pt film is plotted.  The Pt film is attached to a rotating disk at 50mm from rotation axis.
$V_{\rm ISHE}$ is proportional to $B^{2}$ and $\Omega$ when $\omega_{c} \tau \ll 1$. (a) $V_{\rm ISHE}$ is plotted as a function of the magnetic field $B$ at $\Omega=1$kHz.  (b) $V_{\rm ISHE}$ is plotted as a function of the magnetic field $\Omega$ at $B=1$T. }
\label{Vishe}
\end{figure}

In this section, we investigate a detection method of the spin current induced by rotation using the inverse spin Hall effect (ISHE).\cite{Saitoh2006,Ando2011JAP}
The ISHE converts a spin current into an electric voltage by means of the SOI as
\begin{eqnarray}
\E_{\rm ISHE} \propto \v{J}_{s} \times \v{s}
\end{eqnarray}
with the electric field due to the ISHE, $\v{E}_{\rm ISHE}$, the spatial direction of the spin current $\v{J}_{s}$, and the spin polarization vector of the spin current $\v{s}$.
In the Pt film attached to a rotating disk (see Fig. \ref{Vishe}), the electric field $\E_{\rm ISHE}$ is generated in the azimuthal direction since the $z$-polarized spin current $\v{J}_{s}(R)$ is mainly induced in the radial direction when $\omega_{c} \tau \ll 1$.

Our setup of the Pt film is an open circuit, which means that the conventional Hall effect originating from $\E_{\r}$ in Eq. (\ref{Sol:Boltzmann-a}) is not created. In the film, the charge degree of freedom is quickly frozen due to the charge accumulation at the ends of the sample.

In contrast, owing to the nonconservative nature of the spin degree of freedom, spin current is not fully compensated and can flow steadily even in the open circuit.
The compensation of spin current occurs within the spin diffusion length from the ends. We obtain the spin accumulation at the sample ends by solving the spin diffusion equation with a source term originating from the rotationally induced spin current.\cite{Mamoru2011a} The explicit solution of the spin diffusion equation is given in Appendix \ref{Sec:SpinDiffusionEquation} .

The voltage of the ISHE, $V_{\rm ISHE}$, is estimated by\cite{Ando2011JAP}
\begin{eqnarray}
V_{{\rm ISHE}} = \Theta w \rho \langle  J_{s}^{r} \rangle 
\end{eqnarray}
 with the spin Hall angle $\Theta$, sample width $w$, and resistivity $\rho$.
The average of the radial spin current $ \langle  J_{s}^{r} \rangle $ is calculated by
\begin{eqnarray}
\langle  J_{s}^{r} \rangle = \frac{1}{d} \int_{R_{1}}^{R_{2}} J_{s}^{r}dR =\frac{ \omega_{c}^{2}\tau ne \kappa  (R_{1}+R_{2})}{1+ (\omega_{c}\tau )^{2}}. \label{AvJsr}
\end{eqnarray}
From Eq. (\ref{AvJsr}), we obtain the voltage 
\begin{eqnarray}
V_{\rm ISHE} \propto B^{2} \Omega,
\end{eqnarray}
when $\omega_{c}\tau \ll 1$.
This shows that the sign of the voltage changes when the rotation axis is reversed whereas it does not when the magnetic field is reversed (see Fig. \ref{Vishe}).
In the case of the Pt thin film with $w=d=10$mm, $\omega_{c} \tau=0.01$, $\Theta=0.01$ and $\rho=12.8\times 10^{-8} \Omega\cdot$m, 
which is attached to a rotating disk with $R\simeq$ 50mm, the average of the radial spin current is 
\begin{eqnarray}
\langle  J_{s}^{r} \rangle \approx 10^{5} {\rm A/m^{2} }
\end{eqnarray} and
the voltage accross the width of the sample is estimated (see Fig. \ref{Vishe}) as
\begin{eqnarray}
V_{{\rm ISHE}} \approx 1 \mu \mbox{V}.
\end{eqnarray}

\section{Linear acceleration\label{Sec:Linear}}
In this section, we discuss spin current generation by the linear acceleration in the absence of electromagnetic fields.
The Dirac Hamiltonian in a linearly and rotationally accelerating frame without electromagnetic fields was derived by Hehl and Ni:\cite{Hehl1990}
\begin{eqnarray}
H=&&\beta mc^{2} + c \valpha \cdot \p + \frac{1}{2c}[(\va \cdot \r)(\p \cdot \valpha)+(\p \cdot \valpha)(\va \cdot \r)] \nonumber \\
&&+\beta m (\va \cdot \r) - \vOmega \cdot ( \v{L} + \v{S} ),
\end{eqnarray}
with the linear acceleration $\v{a}$ and the angular momentum $\v{L}=\r \times \p$. In the low energy limit of this Hamiltonian up to the order of $1/m^{2}$, one has the Hamiltonian of the Pauli-Schr\"odinger equation of electron's 2-spinor wave function in the accelerating frame:\cite{Hehl1990}
\begin{eqnarray}
H &=& \frac{\p^{2}}{2m} + m \va \cdot \r - \vOmega \cdot (\v{L} + \v{S}) \nonumber \\
&&+ \frac{\hbar}{4mc^{2}} \vsigma \cdot (\va \times \p) \label{H:HN},
\end{eqnarray}
where the rest mass energy and the redshift effect of the kinetic energy\cite{Hehl1990} are neglected. 
We note that the last term, the so-called ``inertial spin-orbit interaction'', does not contain the mechanical rotation because of the absence of the magnetic field.
It should be emphasized that the inertial effect of the linear acceleration on electon can be interpreted as an ``effective electric field''. 
Introducing the ``electric field''
\begin{eqnarray}
\E_{\va} = (m/q) \va ,
\end{eqnarray}
the inertial SOI can be rewritten as
\begin{eqnarray}
H_{\rm S, \va}=\frac{\lambda}{\hbar} \vsigma \cdot (\p \times q\E_{\va}) \label{Hsa}.
\end{eqnarray}
Together with the second term of Eq. (\ref{H:HN}), $m\va \cdot \r = q \E_{\va} \cdot \r$,
the inertial effects of the linear acceleration without rotation on the large spin-orbit interacting system can be analyzed
by the same framework as the conventional Hamiltonian with SOI:
\begin{eqnarray}
H=\frac{\p^{2}}{2m} + U+ \frac{\lambda}{\hbar} \sigma \cdot \left[ \frac{\del U}{\del \r} \times \p \right]
\end{eqnarray}
with the potential $U=q \E_{\va} \cdot \r$. 

The electron velocity in the linearly accelerating frame is given by
\begin{eqnarray}
\dot{\r} = \frac{\p}{m}  + \vv_{\vsigma,\va} 
\end{eqnarray}
where
\begin{eqnarray}
 \vv_{\vsigma,\va} = \frac{e \lambda}{\hbar} \vsigma \times \E_{\va}.
\end{eqnarray}

\begin{figure}[tbp]
\begin{center}
\includegraphics[scale=0.52]{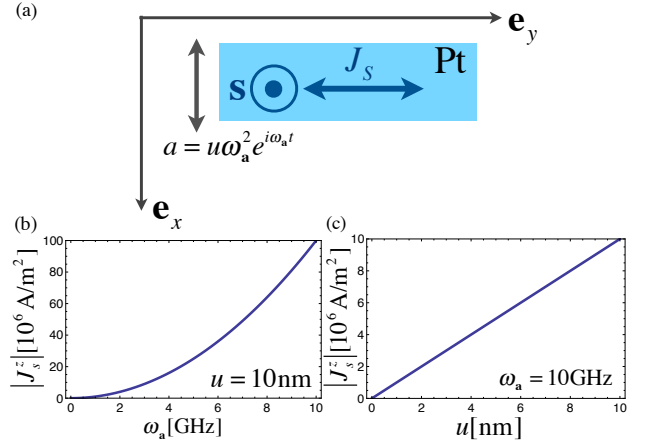}
\end{center}
\caption{Spin current generation in a linearly accelerating frame is shown. (a) When a Pt film is attached to a mechanical resonator and vibrated in $x$-axis, the $z$-polarized ac spin current is created in the $y$-direction.  
(b) The amplitude of $J_{s}^{z}$ is plotted as a function of $\omega_{\v{a}}$ at $u=10$nm. (c) The amplitude is plotted as a function of $u$ at  $\omega_{\v{a}}=10$GHz.}
\label{piezo}
\end{figure}

Let us focus on the inertial effects of linear acceleration. 
The anomalous velocity $\vv_{\vsigma, \va}$ yields the mechanical analogue of the spin Hall effect in a ballistic regime.
The $i-$polarized spin current ($i=\{ x,y,z \}$) generated by the linear acceleration is estimated as 
\begin{eqnarray}
\v{J}_{s}^{i}= en{\rm Tr}[ \sigma_{i} \vv_{\vsigma, \va}] =  \frac{2nem\lambda}{\hbar} \v{s}_{i} \times \va.
\end{eqnarray}
When the acceleration is induced by the harmonic oscillation with the frequency $\omega_{\va}$ and amplitude $u$ in $x$-direction as
\begin{eqnarray}
\va = u \omega_{\va}^{2} e^{i \omega_{\va}t} \ve_{x}
\end{eqnarray}
the $z$-polarized spin current is created in $y$-direction
\begin{eqnarray}
\v{J}^{z}_{s} = \frac{2nem\lambda}{\hbar} u \omega_{\va}^{2}e^{i \omega_{\va}t} \ve_{y}.
\end{eqnarray}
Assuming the Pt film vibrated with $\omega_{\va}= 10$GHz and $u=10$nm, 
the ac spin current is estimated to be $J_{s}^{z}\approx 10^{7} {\rm A/m^{2}}$ (see Fig. \ref{piezo}).
It is a future challenge to probe the ac spin current in the high frequency mechanical resonator.\cite{Mahboob2011}
In such a noninertial system, it is straightforward to extend the results in the ballistic regime to those in the diffusive regime using a well-established framework on the spin Hall effect
by replacing the usual electric field $\E$ with the effective one $\E_{\va}$.

\section{Conclusion\label{Sec:Conclusion}}
In this paper, we have investigated theoretically the generation of spin currents in both rotationally and linearly accelerating systems.
The explicit form of the spin-dependent inertial force acting on electrons in a rotating frame in the presence of electromagnetic fields 
was derived from the generally covariant Dirac equation. 
It has been shown that the force is responsible for the generation of spin currents by mechanical rotation in the first order of  the spin-orbit coupling.
The effect of fluctuation of the rotation axis on the spin current was discussed using the time-dependent part of the force for future experimental analysis.

We have also studied the spin current generation from the mechanical oscillation.
The spin current can be created in a uniformly oscillating conductor with a large spin-orbit coupling because of the inertial spin-orbit interaction originating from the linear acceleration.
We provided a concise interpretation of an inertial effect of the linear acceleration on electron as an effective electric field, 
which allows us to use the conventional theory of the spin Hall effect for describing the spin current generation due to the linear acceleration, by the simple substitution of the effective electric field for an ordinary one.

The framework proposed here offers a new route to study the inertial effects on electron transport phenomena, leading to an innovative combination of microelectromechanical systems (MEMS) and spintronics.




%

%

\begin{acknowledgments}
We are grateful to J. Suzuki, E. M. Chudnovsky, and J. Bogan for valuable comments. 
We thank fruitful discussions and interactions in Spin Caloritronics III organized by the Lorentz Center, and its hospitality. 
This work was supported by a Grant-in-Aid for Scientific Research from MEXT, Japan and the Next Generation Supercomputer Project, Nanoscience Program from MEXT, Japan.

\end{acknowledgments}

\appendix

\section{Metric, vierbein, and spin connection in a rotating frame \label{Sec:Metric}} 

In Sec. \ref{Sec:DiracEquation}, we use the vierbein representation of the Dirac equation.  
The relation of the matrices between curved space-time and flat one is provided by the vierbein (or tetrad) filed $e_{(\alpha)}^{\mu}(x)$, whose indices $\mu$ and $(\alpha)$ labels the curved space-time coordinates and the local flat space-time coordinates, respectively. The vierbein is a local orthonormal base $\{  e^{\mu}_{(\alpha)} \}_{\alpha=0,1,2,3}$ and a kind of square root of the metric tensor $g^{\mu\nu}(x)$: 
\begin{eqnarray}
e^{\mu}_{(\alpha)}(x) \eta^{\alpha\beta} e^{\mu}_{(\beta)} = g^{\mu \nu}(x), \label{vierbein}
\end{eqnarray}
with the Lorentz metric $\eta^{\mu\nu}={\rm diag(-1,1,1,1)}$.
Starting with the metric tensor: 
\begin{eqnarray}
g_{00}=-1+ u^{2},\  g_{0i}= g_{i0}= u_{i},\  g_{ij}=\delta_{ij}, \label{g}
\end{eqnarray}
its inverse tensor is
\begin{eqnarray}
g^{00}=-1,\ g^{0i}=g^{i0}=u_{i},\ g^{ij}=\delta_{ij}-u_{i}u_{j}, \label{inv.g}
\end{eqnarray}
where $u_{i}=(\vOmega \times \r/c)_{i}$ and $u^{2}=u_{1}^{2}+u_{2}^{2}+u_{3}^{2}$.
The vierbein in the frame is obtained from Eq. (\ref{vierbein}) and (\ref{inv.g}):
\begin{eqnarray}
&&e^{0}_{(0)}=1,\ e^{0}_{(j)}=0, \nonumber \\
&&e^{i}_{(0)}=-u_{i},\ e^{i}_{(j)}=\delta^{i}_{j}.
\end{eqnarray}
We also have the inverse of the vierbein $e^{(\alpha)}_{\mu}(x)=g_{\mu\nu}(x)\eta^{\alpha\beta}e^{\nu}_{(\beta)}$:
\begin{eqnarray}
e^{(\alpha)}_{0}= \delta^{\alpha}_{0}+ \eta^{\alpha i}u_{i}, \ e^{(\alpha)}_{i}= \delta^{\alpha}_{i}.
\end{eqnarray}
The Clifford algebra in the curved space-time can be written as $\gamma^{\mu}(x)=g^{\mu\nu}(x)e^{(\alpha)}_{\nu}(x) \bgamma_{\alpha}$
with $\bgamma_{0}=i\beta, \bgamma_{i}=-i\beta \alpha_{i}$.
Thus, we have the algebra in the rotating frame
\begin{subequations}
\begin{eqnarray}
&&\gamma^{0}(x)=-\bgamma_{0}= -i\beta ,\nonumber \\
&&\gamma^{i}(x) = u_{i}\bgamma_{0}+ \bgamma_{i}= i\beta u_{i} - i\beta \alpha_{i},
\end{eqnarray}
\end{subequations}
where
\begin{eqnarray}
 \beta=\mat{I}{O}{O}{-I}, \ 
 \vb{\alpha}=\mat{O}{\vb{\sigma}}{\vb{\sigma}}{O}.
\end{eqnarray}

In the vierbein representation, spin connection $\Gamma_{\mu}(x)$ is expressed as
\begin{eqnarray}
\Gamma_{\mu}(x) = - \frac{1}{4}\bgamma_{\alpha} \bgamma_{\beta} e^{(\alpha)}_{\nu}g^{\nu \lambda} (\del_{\mu} e^{(\beta)}_{\lambda} - \Gamma^{\sigma}_{\mu\lambda} e^{(\beta)}_{\sigma}),
\end{eqnarray}
where the affine connection $\Gamma^{\lambda}_{\mu\nu}$ is defined by
\begin{eqnarray}
\Gamma^{\lambda}_{\mu\nu}=\frac12 g^{\lambda \sigma}(\del_{\nu} g_{\sigma \mu}  + \del_{\mu} g_{\sigma \nu} -\del_{\sigma} g_{\mu \nu}  ).\label{affine}
\end{eqnarray}
Substituting (\ref{g}) into (\ref{affine}),
\begin{eqnarray}
&&\Gamma^{0}_{00}= \Gamma^{0}_{i0}=\Gamma^{0}_{ij}=\Gamma^{i}_{jk}=0, \nonumber \\
&&\Gamma^{i}_{00}= \frac{\epsilon_{ijk} \Omega_{j} u_{k}}{c} + \del_{0} u_{i},\ \Gamma^{i}_{j0}=- \frac{\epsilon_{ijk}\Omega_{k}}{c}.
\end{eqnarray}
From the affine connection, Clifford algebra, and vierbein in the rotating frame, we arrive at the spin(spinor) connection:
\begin{subequations}
\begin{eqnarray}
&&\Gamma_{0}(x)=\frac{\bgamma_{i}\bgamma_{j} \epsilon_{ijk} \Omega_{l}}{4c} = \frac{ i \vb{\Sigma} \cdot \vOmega}{2c}, \\
&&\Gamma_{i}(x)=0.
\end{eqnarray}
\end{subequations}

\section{Foldy-Wouthuysen-Tani transformation\label{Sec:FWTT}}
In Sec. \ref{Sec:DiracEquation}, we derive Pauli-Schr\"odinger equation in a rotating frame by Foldy-Wouthuysen-Tani transformations (FWTTs). The details of the derivation are shown below.
FWTT is a unitary transformation and a block diagonalization method of the Dirac Hamiltonian. First of all, we define the even and odd part of the Hamiltonian as
\begin{eqnarray}
H&=& \beta mc^{2} + \CE + \CO, 
\end{eqnarray}
with
\begin{subequations}
\begin{eqnarray}
\CE &=& qA_{0} - \vb{\Omega} \cdot (\r \times \vb{\pi}+ \vb{\Sigma}),  \\
\CO &=&  c \valpha \cdot \vpi. 
\end{eqnarray}
\end{subequations}
The even part, $\CE$, is the block diagonal part of the Hamiltonian whereas the odd part, $\CO$, is the block off-diagonal part.
FWTT is defined by
\begin{eqnarray}
H'=UHU^{\dagger} - U \del_{t} U^{\dagger}  \label{FWTT1}
\end{eqnarray}
with
\begin{eqnarray}
U=\exp \left( -\frac{i\beta \CO}{2mc^{2}} \right)=\exp \left( -\frac{i \beta c\vb{\alpha} \cdot \vb{\pi}}{2mc^{2}} \right).
\end{eqnarray}
A low energy expansion of the Hamiltonian $H'$ into an exponential series of $1/m$ gives a systematic expansion in a way that odd parts of $H'$ vanish in any order.
Up to the order of $1/m^{2}$, Eq. (\ref{FWTT1}) is reduced to
\begin{eqnarray}
H'=&& \beta \left[ mc^{2}+ \frac{\CO^{2} }{2mc^{2}}  \right]  
+\CE - \frac{1}{8m^{2}c^{4}} \left[\CO, \left[ \CO, \CE \right]+ i\hbar \dot{\CO} \right],\nonumber\\ \label{FWTT2}
\end{eqnarray}
Using the relation
\begin{subequations}
\begin{eqnarray}
&&\alpha_{i} \alpha_{j} = \delta_{ij} + i \epsilon_{ijk}\sigma_{k}, \\
&& [ \pi _{i}, \pi _{j} ] = i \hbar q \epsilon_{ijk} B_{k}
\end{eqnarray}
\end{subequations}
the second term of Eq. (\ref{FWTT2}) becomes
\begin{eqnarray}
\frac{\beta \CO^{2}}{2mc^{2}}=\frac{\beta \alpha_{i} \alpha_{j} \pi_{i} \pi_{j}}{2m}=\beta \left( \frac{\vpi^{2}}{2m}  -\frac{q\hbar}{2m} \vsigma \cdot \v{B} \right). \label{KinZeeman}
\end{eqnarray}
Next, we focus on $[\CO,[\CO, \CE]+i\hbar \dot{\CO}]$.
\begin{eqnarray}
[\CO,\CE]&+&i\hbar \dot{\CO} \nonumber\\
&=& [c \valpha \cdot \vpi, -qA_{0}] + c\valpha \cdot (-iq \hbar) \del_{t} \v{A} \nonumber \\
&&+ [c \valpha \cdot \vpi, - \vOmega \cdot (\vSigma + \r \times \vpi) ] \nonumber \\
&=& -i \hbar c q \valpha \cdot \left( - \nabla A_{0} + \del_{t} \v{A}  \right)\nonumber \\
&& - i \hbar c \valpha \cdot (\vOmega \times \vpi)+ i \hbar c \valpha \cdot (\vOmega \times \vpi) \nonumber \\
&& - i \hbar c q (\vOmega \times \r) \times \B \nonumber \\
&=& i \hbar c \valpha \cdot q\v{E}'
\end{eqnarray}
with
\begin{eqnarray}
\v{E}'= - \nabla A_{0} + \del_{t} \v{A} + (\vOmega \times \r) \times \B.
\end{eqnarray}
Thus, we have
\begin{eqnarray}
- \frac{1}{8m^{2}c^{4}} &&\left[\CO , \left[ \CO, \CE \right]+ i\hbar \dot{\CO} \right] \nonumber \\
&&=- \frac{1}{8m^{2}c^{4}} \left[ c \valpha \cdot \vpi, -i \hbar c \valpha \cdot q \v{E}'  \right] \nonumber \\
&& = \frac{i\hbar}{8 m^{2}c^{2}} \alpha_{i} \alpha_{j} [\pi_{i} , qE'_{j}] \nonumber \\
&& = \frac{q\hbar^{2}}{8 m^{2}c^{2}} {\rm div} \E'  \nonumber \\
&&\ + \frac{q\hbar}{8 m^{2}c^{2}} \vsigma \cdot ( \vpi \times  \E' - \E' \times \vpi). \label{SOI-Erot}
\end{eqnarray}
From Eqs. (\ref{KinZeeman}) and (\ref{SOI-Erot}), we obtain the Hamiltonian $H_{\rm PR}$ in Sec. \ref{Sec:PauliRot}, by neglecting the rest energy term.

\section{Electromagnetic fields in a rotating frame\label{Sec:EMrot}}
In this Appendix, we give a brief review of a general relativistic transformation of electromagnetic fields.

Electromagnetic fields in the rotating frame are related to those in the rest frame as
\begin{subequations}\label{RotTr}
\begin{eqnarray}
\E' &=& \E + (\vOmega \times \r) \times \B, \label{RotTr-a}\\
\B' &=& \B,\label{RotTr-b}
\end{eqnarray}
\end{subequations}
when the rotation velocity is much less than the speed of light, $|\vOmega \times \r | \ll c$, where $\E'$ and $\B'$ are electromagnetic fields in the rotating frame.

Equations (\ref{RotTr}) are not Lorentz transformation in special relativity but a general coordinate transformation in general relativity. 
The Lorentz transformations of the electromagnetic fields are written as\cite{LandauFields} 
\begin{subequations}\label{LorentzTr}
\begin{eqnarray}
\E''/c &=& \gamma (\E/c +  \vbeta \times \B) -\frac{\gamma^{2}}{\gamma + 1} (\vbeta \cdot\E/c) \vbeta  \label{LorentzTr-a}\\
\B'' &=& \gamma (\B - \vbeta \times \E/c) - \frac{\gamma^{2}}{\gamma + 1}(\vbeta \cdot \B) \vbeta, \label{LorentzTr-b}
\end{eqnarray}
\end{subequations}
with
\begin{eqnarray}
\gamma &=& \frac{1}{\sqrt{1-\vbeta^{2}}}, \ \vbeta = \frac{\vv_{0}}{c}.
\end{eqnarray}
Here $\E''$ and $\B''$ are the electromagnetic fields in the inertial frame which has a uniform velocity $\vv_0$ relative to the rest frame. Replacing $\vv_0$ with $\vv(x)=\vOmega \times \r$ in Eqs. (\ref{LorentzTr}) can never reproduce the correct transformation between the rotating frame and the rest frame (\ref{RotTr}).
The special relativistic transformations (\ref{LorentzTr}) have an apparent symmetry for $\E/c$ and $\B$ whereas the relations (\ref{RotTr}) do not. 
Such an asymmetry in (\ref{RotTr}) originates from the space-time asymmetry of a general coordinate transformation which relates physical quantities in a rotating frame to those in a rest frame.

When the rotation axis is parallel to the $z$-axis, the transformation between the rest frame and the rotating frame is expressed as
\begin{eqnarray}
x'^{\mu} = L^{\mu}_{\nu} x^{\nu}\label{TrL}
\end{eqnarray}
with
\begin{eqnarray}
 L^{\mu}_{\nu}=
\left(\begin{array}{cccc}
1 & 0 & 0 & 0 \\
0 & \cos \Omega t & \sin \Omega t & 0 \\
0 & -\sin \Omega t & \cos \Omega t & 0 \\
0 & 0 & 0 & 1\end{array}\right),
\end{eqnarray}
where rotating coordinates carry the prime. 
Equation (\ref{TrL}) leads to a general coordinate transformation\cite{Ridgely1999,LandauFields}
\begin{eqnarray}
R=\frac{\del x^{\alpha}}{\del x'^{\beta}} =
\left(\begin{array}{cccc}
1 & 0 & 0 & 0 \\
\Omega y/c & \cos \Omega t & \sin \Omega t & 0 \\
-\Omega x/c & \sin \Omega t & \cos \Omega t & 0 \\
0 & 0 & 0 & 1 
\end{array}\right).\label{TrMCRF} 
\end{eqnarray} 
According to the principle of general covariance, electromagnetic fields are components of a second rank tensor in general coordinate transformations. 
Electromagnetic tensors in the rest frame $F$ and those in the rotating frame $F'$ are related as:
\begin{eqnarray}
F'=L R^{T} F R L^{T}, \label{RTFR}
\end{eqnarray} 
in which $T$ denotes the transpose matrix, and
\begin{eqnarray}
F=
\left(\begin{array}{cccc}
0 & -E_x/c & -E_y/c & -E_z/c \\
E_x/c & 0 & B_z & -B_y \\
E_y/c & -B_z & 0 & B_x \\
E_z/c & B_y & -B_{x} & 0
\end{array}\right).\label{Fmunu}
\end{eqnarray}
Equations (\ref{RTFR}) and (\ref{Fmunu}) lead to
\begin{subequations}
\begin{eqnarray}
E'_{x} &=& E_{x }+ x\Omega  B_{z} \\
E'_{y} &=& E_{y} + y\Omega  B_{z} \\
E'_{z} &=& E_{z}                          - \Omega  (x B_{x}+y B_{y}) \\
B'_{x} &=& B_{x }  \\ 
B'_{y} &=& B_{y}  \\ 
B'_{z} &=&B_{z}.\label{EBOmegaT}
\end{eqnarray}
\end{subequations}
Thus, we obtain the relations (\ref{RotTr}) in Sec. \ref{SubSec:2nd}. 
The relations shown in this section hold for $|\vOmega \times \r | \ll c$. Generalized relations for electromagnetic fields in more rapidly rotating frame should include a Lorentz factor $\gamma=1/\sqrt{1-(\vOmega \times \r/c)^{2}}$.\cite{Ridgely1999}
We omit discussion on Maxwell's equations in a rotating frame. The subject is originally studied by means of generally covariant Maxwell's equations\cite{Schiff1939}.

\section{Solution of the spin diffusion equation in a rotating frame\label{Sec:SpinDiffusionEquation}}
In this Appendix, we discuss the spin diffusion equation, to clarify the spin-dependent transport property of spin current  in a rotating frame. 
In the non-equilibrium steady state, spin current in the rotating frame consists of the diffusive current $J_{s}^{\rm diff}$ and the drift currnt $J_{s}^{\rm rot}$:
\begin{eqnarray}
J_{s}^{\rm tot}= J_{s}^{\rm diff} + J_{s}^{\rm rot}
\end{eqnarray}
with
\begin{eqnarray}
J_{s}^{\rm diff} = \frac{1}{e \rho}\nabla \delta \mu. 
\end{eqnarray}
Here, $\delta \mu$ is the spin accumulation.
The drift part, which can be regarded as a spin source term, is given by the mechanically induced spin current (\ref{Js:r-phi}) in Sec. VI.
The spin diffusion equation with the source term is written as
\begin{eqnarray}
\nabla^{2} \delta \mu = \frac{1}{\lambda_{s}^{2}} \delta \mu - e  \rho\,  \mbox{div} J_{s} \label{SpinDiff}
\end{eqnarray}
with the spin diffusion length $\lambda_{s}$.
When both rotation axis and magnetic field are parallel to the $z$-axis, $J_{s}^{\rm rot}$ is mainly induced in the radial direction in the Pt film as discussed in the text. Thus, we solve this equation in the cylindrical coordinate. 
The radial component of the equation is reduced to
\begin{eqnarray}
\frac{1}{ r }\frac{\del }{\del r} \left(  r \frac{\del \delta \mu}{\del r} \right) = \frac{\delta \mu}{\lambda_{s}^{2}}  - \frac{e\rho}{r} \frac{\del}{\del r}\left( r\cdot  \frac{2ne\kappa \tau \omega_{c}^{2} r}{1+(\tau \omega_{c})^{2}}  \right).
\end{eqnarray}
Using parameters 
\begin{eqnarray}
a=\frac{1}{\lambda_{s}}, \ b= 2e\rho A, \ A=\frac{2ne\kappa \tau \omega_{c}^{2} }{1+(\tau \omega_{c})^{2}},
\end{eqnarray}
the analytical solution can be written as
\begin{eqnarray}
\delta \mu(r) = \frac{b}{a^{2}} + c_{1} J_{0}(a r) + c_{2} K_{0}(a r),
\end{eqnarray}
where $J_{n}, K_{n}$ are the modified Bessel function of the first and second kind, and the constants $c_{1}$ and $c_{2}$ are to be determined by the boundary condition:
\begin{eqnarray}
J_{s}^{\rm tot}(R_{i})=0, \ (i=1,2)
\end{eqnarray}
at the ends of the Pt, $r=R_{1}, R_{2}$. In this condition, we have
\begin{subequations}
\begin{eqnarray}
&&c_{1} = c_{0}[ R_{2}J_{1}(aR_{1})+ R_{1} J_{1}(aR_{2})]   \\
&&c_{2} = c_{0} [-R_{2} K_{1}(aR_{1}) + R_{1} K_{1}(aR_{2})]
\end{eqnarray}
\end{subequations}
with
\begin{eqnarray}
c_{0}= \frac{b}{a [J_{1}(aR_{2}) K_{1}(aR_{1}) - J_{1}(aR_{1}) K_{1}(aR_{2})]}.
\end{eqnarray}
Putting $d=100$nm, $R_{1}=50$mm, $\Omega=1$kHz, $B=1$T, $\lambda_{s}=14$nm, and $\rho=12.8 \times 10^{-8}\Omega\cdot$m for Pt,\cite{Steenwyk1997}  we obtain the spin accumulation near the ends, $\delta \mu(R_{1}) \approx 0.05$nV and $\delta \mu(R_{2}) \approx -0.05$nV.

\end{document}